\documentclass[reprint, superscriptaddress, amsmath, amssymb, aps, pre]{revtex4-2}

\usepackage{graphicx}
\usepackage{dcolumn}
\usepackage{bm}
\usepackage{wasysym}

\usepackage[dvipsnames]{xcolor}

\usepackage{hyperref}
\hypersetup{colorlinks = true, 
urlcolor  = BrickRed, 
linkcolor = NavyBlue, 
citecolor = ForestGreen}

\usepackage{amsthm}
\usepackage{amsmath}
\usepackage{amssymb}

\begin{document}

\title{Inverse linear vs. exponential scaling of work penalty in finite-time bit reset }
\author{Yi-Zheng Zhen}
\email{yizheng@ustc.edu.cn}
\affiliation{Hefei National Laboratory for Physical Sciences at Microscale and Department of Modern Physics,
University of Science and Technology of China, Hefei, 230026, China}
\author{Dario Egloff}
 \affiliation{Institute of Theoretical Physics, Technical University Dresden, D-01062 Dresden, Germany}
 \affiliation{Max  Planck  Institute  for  the  Physics  of  Complex  Systems, N{\"o}thnitzer  Strasse  38,  01187  Dresden,  Germany}
\author{Kavan Modi}
 \affiliation{School of Physics and Astronomy, Monash University, Clayton, Victoria 3800, Australia}
\author{Oscar Dahlsten}
\email{dahlsten@sustech.edu.cn}
\affiliation{Shenzhen Institute for Quantum Science and Engineering and Department~of~Physics,Southern University of Science and Technology, Shenzhen 518055, China}
\date{\today}

\begin{abstract}
Bit reset is a basic operation in irreversible computing. This costs work and dissipates energy in the computer, creating a limit on speeds and energy efficiency of future irreversible computers. It was recently shown in Ref. \href{https://doi.org/10.1103/PhysRevLett.127.190602}{[Phys. Rev. Lett. 127, 190602 (2021)]} that for a finite-time reset protocol, the additional work on top of the quasistatic protocol can always be minimized by considering a two-level system, and then be lower bounded through a thermodynamical speed limit. An important question is to understand under what protocol parameters, including bit reset error and maximum energy shift, this penalty decreases exponentially vs inverse linearly in the protocol time. Here we provide several analytical results to address this question, as well as numerical simulations of specific examples of protocols. 

\end{abstract}

\maketitle

\section{Introduction}\label{sec:introduction}

It has been a long-term effort to reduce the energy consumption of computing.
As an elementary operation in irreversible computation, bit reset initializes unknown logical bits to certain states.
Understanding the work cost to reset a bit is important when dealing with the energetic cost of irreversible computation~\cite{LentOPS19}.
Indeed, not only does the work cost contribute to the overall energy cost of the computation, but also the excess work dissipates as heat, which limits the number of computations per second. Landauer's principle~\cite{Szilard29, Landauer61, Bennett82} imposes a fundamental limit to such energy cost: the erasure of one bit of information requires at least $k_B T\ln2$ amount of energy, where $k_B$ is the Boltzmann constant and $T$ is the temperature of the environment. This limit has been extrapolated based on current trends to be reached around 2035~\cite{Frank05}. The principle is moreover of fundamental significance to the foundation of information thermodynamics~\cite{piechocinska_information_2000, vedral_landauers_2000,  plenio_physics_2001, dillenschneider_memory_2009,  delRioARDV11, Aberg11, DahlstenRRV11, EgloffDRV15, reeb_improved_2014, goold_nonequilibrium_2015, lorenzo_landauers_2015, taranto_emergence_2018, BoydMC18, ScandiM19, AbiusoMLS20, timpanaro_landauers_2020, miller_quantum_2020, riechers_impossibility_2021, Taranto+21, VanVuS2021}, and has been tested in various experiments~\cite{BerutAPCDL12, Orlov12, Jun14, koskiMPA14, Peterson16, Gaudenzi18, Yan18}.

Since Landauer's principle assumes quasistatic processes to obtain the limit of energy cost, it is necessary to ask how this limit can be improved in realistic processes.
For a finite-time protocol of bit reset, the crucial question is how to characterize a correction term added to the Landauer limit.
Developments have been made towards solving this problem, by discussing the extra energy cost (or heat dissipation) of bit reset in different systems and scenarios.
For instance, in the Langevin dynamics  case of colloidal particles trapped in a double-well potential, it is shown that optimal reset protocols~\cite{AurellMM11, ZulkowskiD14} yield a correction of heat dissipation proportional to $1/\tau$~\cite{ProesmansEB20, ProesmansEB20pre}, where $\tau$ is the finite time for the reset.
Using a two-level system, it was shown that there exist efficient reset protocols where the work cost drops faster when certain assumptions can be made~\cite{BrowneGDV14}.

Recently, the above two results were further unified and generalized in the framework of stochastic thermodynamics~\cite{ZhenEMD21}.
It was shown that the work penalty, defined as the difference between the actual work cost and quasistatic work, 
is the summation of two terms: the first is the change of relative entropy of the system state and the thermal state, and the second is the entropy production for the entire process.
For bit reset, the entropy production cannot escape $1/\tau$ time scaling~\cite{ZhenEMD21}, guaranteed by the speed limit~\cite{ShiraishiFS18}, and that there exists a region of small $\tau$ where the relative entropy change dominates the work cost and decays exponentially with~$\tau$~\cite{ZhenEMD21}.

While this result applies to a large class of reset protocols,
in implementations, one still needs to design concrete protocols to tailor the work penalty, such that specific requirements (e.g., limitations on energy input, reset time, reset error, or thermalisation rate) can be fulfilled.
Therefore, it is necessary to investigate the trade-off relations between protocol parameters for typical classes of reset protocols.

In this paper, we therefore give a precise analysis of the work penalty for two kinds of bit reset protocols. We focus on two concrete protocols where two-level systems undergo  thermalization that respects detailed balance: a protocol with discrete shifts of energy levels and one with continuous driving.
We study both in two scenarios of fixed maximal energy level and fixed reset error.
We show that for both cases, the work penalty decays exponentially in $\tau$ for small $\tau$ and inverse linearly for large 
$\tau$. We give, for one explicit protocol, a simple sufficient condition on $\tau$ for the exponential scaling to dominate.
We further show analytically and numerically how the work penalty and reset error changes with the allowed maximal energy and the allowed reset error.
Our work gives a more comprehensive understanding of bit reset protocols, and helps to improve the design of bit reset protocols with system-dependent conditions.

\section{Work penalty of bit reset}\label{sec:preliminaries}

\begin{figure}[htp]
\centering
\includegraphics[trim={2.2cm 1.9cm 1.2cm 2.5cm}, clip, width=1.0\columnwidth]{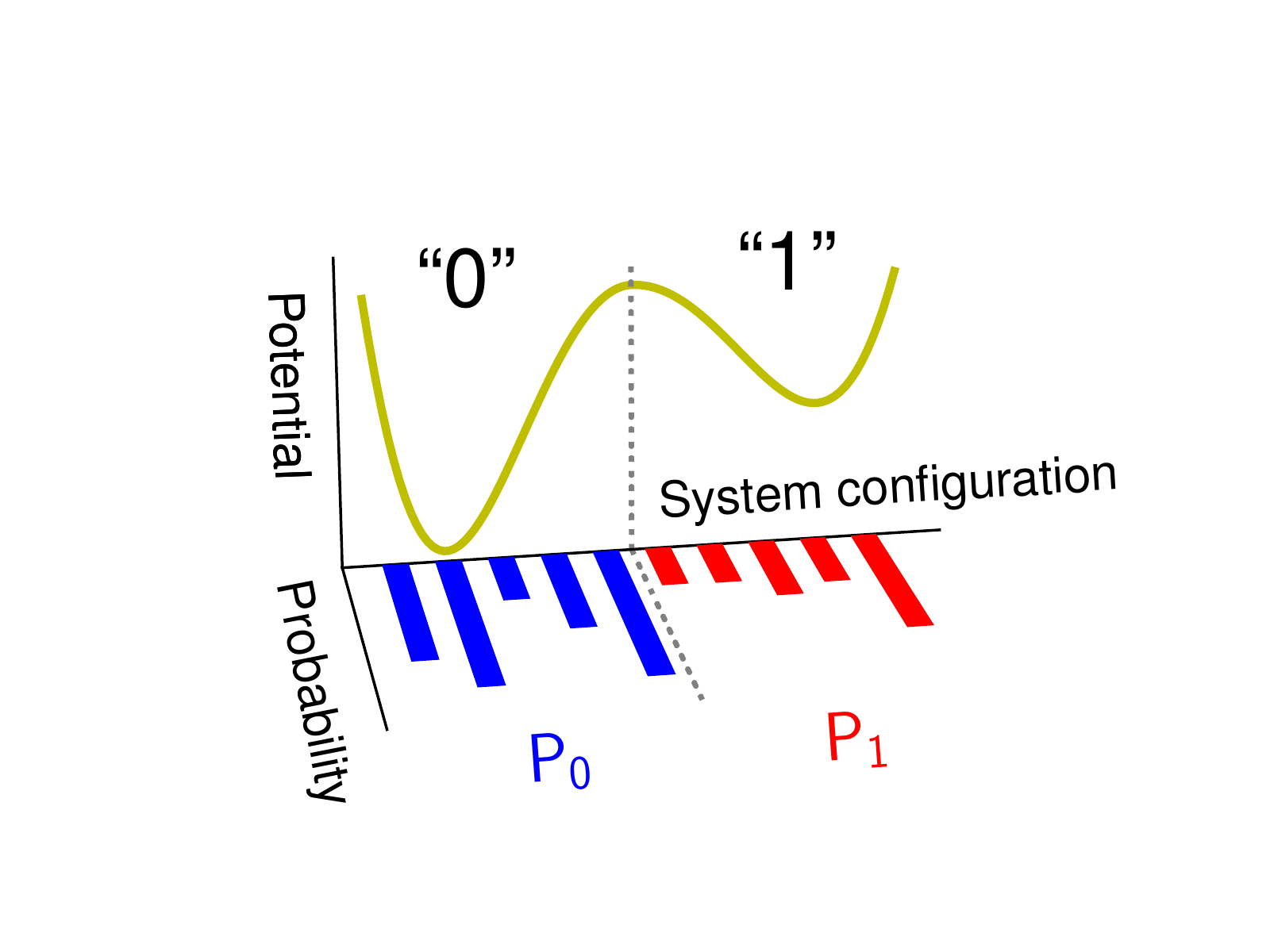}
\caption{\label{fig:model} An example of coarse-graining a multi-level system to a bit system. A colloidal particle immersed in a double-well potential (yellow curve) has multiple microstates (configurations). These microstates occur with certain probabilities (blue and red bars), and are grouped into two bit states ``0'' and ``1'', with probabilities $P_0$ and $P_1$, respectively.}
\end{figure}

The bit is the basic physical ingredient for information processing.
It has two logical states ``0'' and ``1''.
In practice, any physical system, whether  discrete or continuous, can be used as a bit through coarse graining~\cite{Esposito12}, i.e., fine-grained states are grouped into two coarse-grained states representing the logical states of a bit. For instance, a colloidal particle in a double-well potential can be regarded as a bit, with the left well representing ``0'' and the right well representing ``1'' (see Fig.~\ref{fig:model} and Refs.~\cite{Landauer61,Bennett82,AurellMM11,BerutAPCDL12,Orlov12,ZulkowskiD14,ProesmansEB20,ProesmansEB20pre}).

Denote probabilities of the bit system in states ``0'' and ``1'' as $P_0$ and $P_1$, respectively. It has been proved that the work penalty of bit reset can be minimized by considering an effective two-level system~\cite{ZhenEMD21}, therefore, here we only consider the two-level system.
For convenience, we write the state of the  system at time $t$ as $P(t)=[P_0(t),P_1(t)]$.
The system is assumed to be in contact with a heat bath at a fixed temperature $T$.
The system is controllable via varying the energies $E_0(t)$ and $E_1(t)$ associated with states ``0'' and ``1'', respectively.
Let the thermal states, which depend on $E(t)$, be $\gamma(t)=[\gamma_0(t),\gamma_1(t)]$ with $\gamma_a(t)=\exp[-\beta E_a(t)]/Z(t)$.
Here, 
$Z(t)=\sum_a\exp[-\beta E_a(t)]$ is the partition function.

We assume there is a Markovian thermalisation process, consistent with the stochastic thermodynamics framework. Any non-Markovian dynamics can be made Markovian by adding the necessary historical record to the microstate description~\cite{BreuerP07,MilzM21}. The master equation of the bit evolution can, assuming detailed balance, 
always be described by the partial-swap thermalization towards the equilibrium state~\cite{ScaraniZSGB02,BrowneGDV14,ZhenEMD21}, i.e. 
\begin{equation}\label{eq:partial-swap-model}
\frac{d}{dt}P(t) = \mu \left[\gamma(t) - P (t) \right].
\end{equation}
Here, for gaining physical insight of the model, we have assumed that the the swap rate $\mu$ is time-independent and that the equilibrium state is the thermal state.
In general, the swap rate and the equilibrium may depend on the hidden degrees of freedom.
The meaning of partial swap can be seen from taking a small time duration $\delta t$, when a proportion of $\mu(t)\delta t$ of system state $P(t)$ is swapped with the thermal state $\gamma(t)$.

Bit reset is the process of erasing an unknown bit, i.e., driving the bit from the maximally unknown state $P=[1/2, 1/2]$ to a definite logical state, say ``0''.
Without losing generality, a bit reset protocol can then be processed as follows.
Initially, one set $E_1(t=0) = E_0(t=0)=0$ and has let the system sufficiently thermalize such that $P(t=0)=\gamma(t=0)=[1/2,1/2]$.
Then, fix $E_0(t)$ for all the times while increases $E_1(t)$ until a final time $t=\tau$.
During the energy shift, $P_1(t)$ decreases as a result of thermalization described by Eq.~\eqref{eq:partial-swap-model}.
Finally, at time $\tau$, the protocol ends and the system evolves to state $P(t=\tau)$.
Denote $\epsilon = P_1(\tau)$ as the reset error.
One can immediately conclude that $\epsilon$ approaches zero if $E_1(\tau)=E_{\max}$ is sufficiently large and $\tau$ is sufficiently long.

The work cost associated with the above protocols depends on how $E_1(t)$ is lifted, i.e.
\begin{equation}\label{eq:work-cost}
 W (\tau) = \int_0^{\tau}dt P_1 (t) \frac{dE_1}{dt} (t).
\end{equation}
The $W(\tau)$ is fundamentally restricted by Landauer's principle when quasistatic protocols are considered.
For such protocols, the system state is exactly the thermal state at each energy configuration, such that the quasistatic work cost is $W_{\rm qs}(E_{\max})=\int_0^{E_{\max}}\gamma_1(E_1)dE_1=k_B T\ln[2/(1+e^{-\beta E_{\max}})]$.
One can further reach Landauer's limit $k_B T\ln2$ \cite{Landauer61} by letting $E_{\max}\rightarrow\infty$.

In a realistic situation, however, not only is the reset protocol far from quasistatic, but also a non-vanishing reset error exists.
We therefore look at the additional work cost in the finite-time scenario, i.e., the work penalty~\cite{BrowneGDV14}, defined by
\begin{equation}\label{eq:work-penalty-def}
 W_{\rm pn}(\tau) = W(\tau)-W_{\rm qs}(E_{\max}).
\end{equation}
The work penalty can be equivalently written as the sum of two terms (see for example Ref.~\cite{EspositoB11, ZhenEMD21})
\begin{equation}\label{eq:work-penalty-sum}
\beta W_{\rm pn}(\tau) = D\left[P(\tau)\|\gamma(\tau)\right] + \Sigma(\tau),
\end{equation}
where $ D[P(\tau)\|\gamma(\tau)] = \sum_a P_a(\tau)\ln [P_a(\tau)/\gamma(\tau)]$ is the relative entropy between $P(\tau)$ and $\gamma(\tau)$, and $\Sigma(\tau)$ is the entropy production. Moreover, the work penalty can be lower bounded by~\cite{ZhenEMD21}
\begin{equation}\label{eq:work-penalty-bound}
\beta W_{\rm pn}\left(\tau\right)\geqslant D_{\epsilon}(\tau) + \frac{(1-2\epsilon)^2}{\mu \tau},
\end{equation}
where $D_\epsilon(\tau)=D[P(\tau)\|\gamma(\tau)]$.

This bound is tight when the protocol is quasistatic~\cite{ZhenEMD21}.
It is clear to see that whilst a part of $W_{\rm pn}(\tau)$ cannot drop faster than $1/\tau$ scaling, if the  $D_{\epsilon}(\tau)$ term dominates in a certain $\tau$ regime,  $W_{\rm pn}(\tau)$ would effectively scale however $D_{\epsilon}(\tau)$ does~\cite{ZhenEMD21}. We now look at the scaling and trade-offs between the relevant quantities, including the work penalty, reset error, and finite times, in specific models. 

\section{Discrete-shifting protocol}\label{sec:const-shifting-discrete}

As a specific, yet simple, protocol to study the variation of the work penalty with protocol parameters, we consider the protocol discussed in Refs.~\cite{BrowneGDV14} and~\cite{ZhenEMD21}.
In this protocol, 
let the initial time be 0 and the final time be $\tau$, which is divided into N equal size steps. At the start of each step the energy is changed by $\mathcal{E} = E_{\rm max} / N$ (we assumed initial energy to 0). Thus in $k$-th step energy goes from $(k-1)\mathcal{E}$ to  $k\mathcal{E}$. We assume this change occurs instantaneously. Followed by the change in energy, the system undergoes thermalization, which takes time $\tau/N$. Therefore, the energy level in the $k$-th step, is
\begin{equation}\label{eq:const-shifting-energies}
E_1^k = k \mathcal{E}, \quad k=1,\dots,N.
\end{equation}
The instantaneous energy change additionally assumes that the interaction between the system and work medium is much larger than the interaction between the system and the heat bath.
According to the partial swap model, $P_1(t=k\tau/N)$, denoted as $P_1^k$, can be derived by solving Eq.~\eqref{eq:partial-swap-model} as
\begin{equation}\label{eq:const-shifting-P1}
P^{k}_1 =e^{-\mu\tau/N}P^{k-1}_1+\left(1-e^{-\mu\tau/N}\right)\gamma^{k}_1.
\end{equation}
Consequently, one can obtain the work cost and work penalty exactly, albeit not in a simple form, from  Eqs.~\eqref{eq:work-penalty-def} and~\eqref{eq:const-shifting-P1}.

We show that, for general protocols on a two-level system, $D_\epsilon(\tau)$ cannot drop faster than an exponential scaling, as it is lower bounded by
\begin{align}
D\left[P\left(\tau\right)\|\gamma\left(\tau\right)\right]
\geqslant \max & \left\{e^{-2\mu t}G_{1}-e^{-\mu t}\left(1-e^{-\mu t}\right)G_{2},\right.\nonumber\\
&\qquad\left.0\right\},\label{eq:relat-entropy-lower-bound}
\end{align}
where $G_{1}=\ln\left[\left(1+\cosh\left(\beta E_{\max}\right)\right)/2\right]/2$
and $G_{2}=\beta\left[W_{{\rm pn}}\left(\tau\right)-\epsilon E_{\max}\right]$.
We give the derivation of Eq.~\eqref{eq:relat-entropy-lower-bound} in Appendix~\ref{subsec:app-relat-entropy-bound}.

In the other direction, $D_\epsilon(\tau)$ is upper bounded by an exponential function~\cite{ZhenEMD21},
\begin{align}
D_{\epsilon}(\tau)&\leqslant e^{-\mu \tau/N}D\left[\gamma(0)\|\gamma(\tau)\right]\\
&=e^{-\mu \tau/N}\beta\left[\frac{E_{\max}}{2}-W_{\rm qs}(E_{\max})\right].\label{eq:const-shifting-relent-upper-discrete}
\end{align}
Together with Eq.~\eqref{eq:relat-entropy-lower-bound}, we can conclude that $D_{\epsilon}(\tau)$ drops exponentially with $\tau$.
Moreover, as $\beta W_{\rm pn}(\tau)$ is the sum of  $D_{\epsilon}(\tau)$ and $\Sigma(\tau)$, and the speed limit approach guarantees that $\Sigma(\tau)\geqslant(1-2\epsilon)^2/(\mu\tau)$, we conclude that $W_{\rm pn}(\tau)$ may have an exponential scaling or an inverse-linear scaling, dependent on which term dominates.

We further derive a sufficient condition such that $W_{\rm pn}(\tau)$ is dominated by $\Sigma(\tau)$, namely
\begin{equation}\label{eq:const-shifting-sigma-dominates}
\tau\geqslant\frac{N}{2\mu}\ln2,
\end{equation}
and give the proof in Appendix~\ref{subsec:app-const-shifting-sigma-dominates}.

We shall also compare the expressions here with upper and lower bounds on $W_{\rm pn}(\tau)$ which were derived in Ref.~\cite{ZhenEMD21} using techniques from Ref.~\cite{BrowneGDV14}:
\begin{align}
W_{\rm pn}(\tau) & \geqslant e^{-\mu\tau/N}\left[\frac{1}{2}-\frac{1}{1+e^{\beta\left(E_{\max}-{\cal E}\right)}}\right]{\cal E},\label{eq:const-shifting-L}\\
W_{\rm pn}(\tau) & \leqslant \frac{e^{\beta{\cal E}}-1}{2\beta}+e^{-\mu\tau/N}\left[\frac{E_{\max}}{2}-W_{\rm qs}(\tau)\right].\label{eq:const-shifting-U}
\end{align}

In the following, we study the performance of these bounds with different parameters $N$, $\epsilon$, $E_{\max}$, and $\mu$.
We particularly discuss two cases: the case of fixed $E_{\max}$ and the case of fixed $\epsilon$. These are two important parameters that could be varied as part of the protocol design. 
$E_{\max}$ can be heavily constrained for given hardware. $\epsilon$ is a key performance indicator: the bit reset error probability. 
We break the task of considering all combinations of these two parameters into two parts by considering one parameter fixed and the other as free.  

\subsection{Case of fixed \texorpdfstring{$E_{\max}$}{Emax} }\label{subsec:const-shifting-fixed-emax}

\begin{figure}[htp]
\centering
\includegraphics[width=.9\columnwidth]{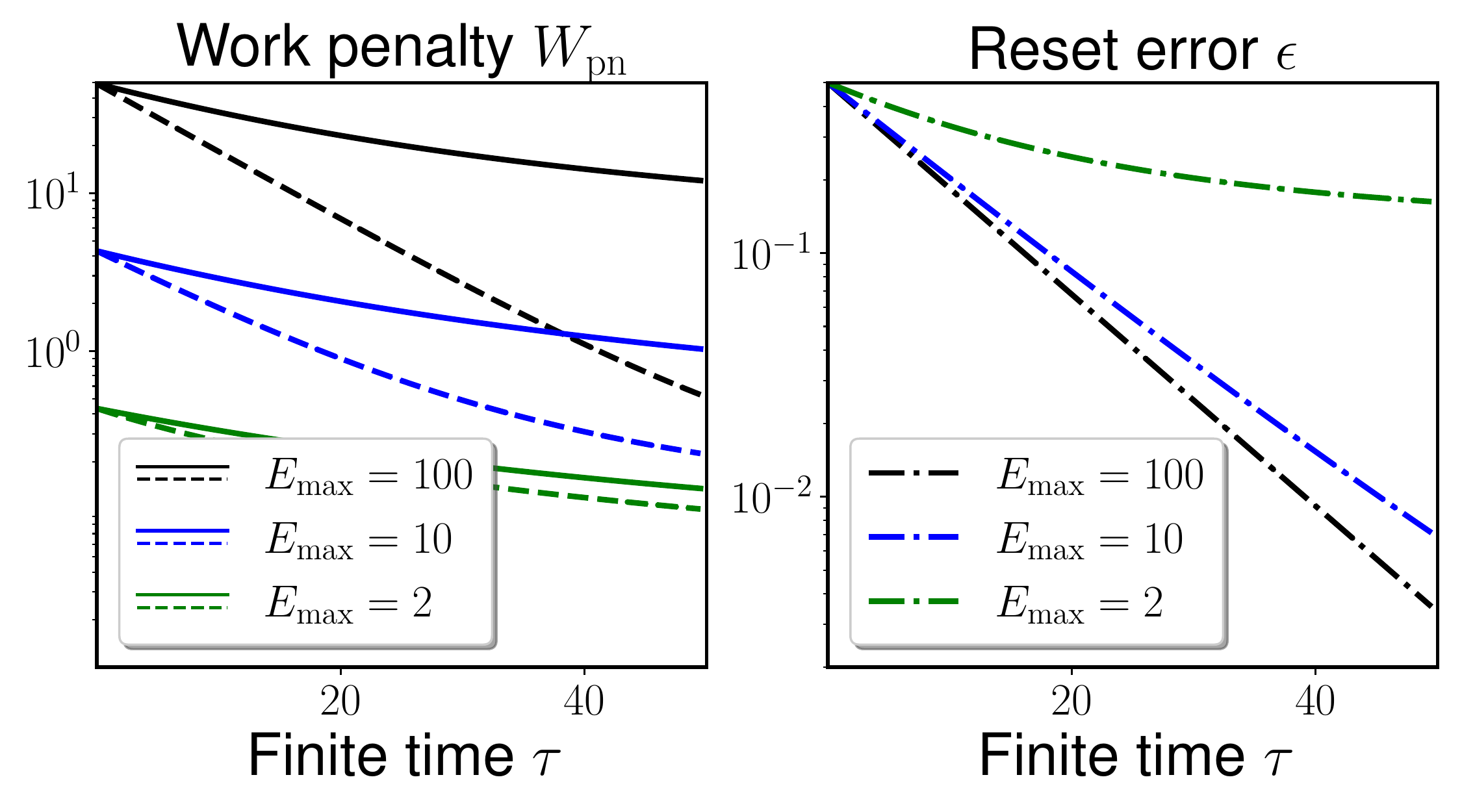}
\caption{\label{fig:Fig_fixE_EmaxVsEps}%
The discrete-shifting protocol with fixed maximal energy $E_{\max}$. The solid lines, dashed lines and dashdotted lines represent the work penalty $W_{\rm pn}(\tau)$, the work penalty bound in Eq.~\eqref{eq:work-penalty-bound} and the reset error $\epsilon$, respectively. $N=10$, $\mu=0.1$ and $\beta=1$ have been chosen for the simulation.}
\end{figure}

\begin{figure}[htp]
\centering
\includegraphics[width=1.0\columnwidth]{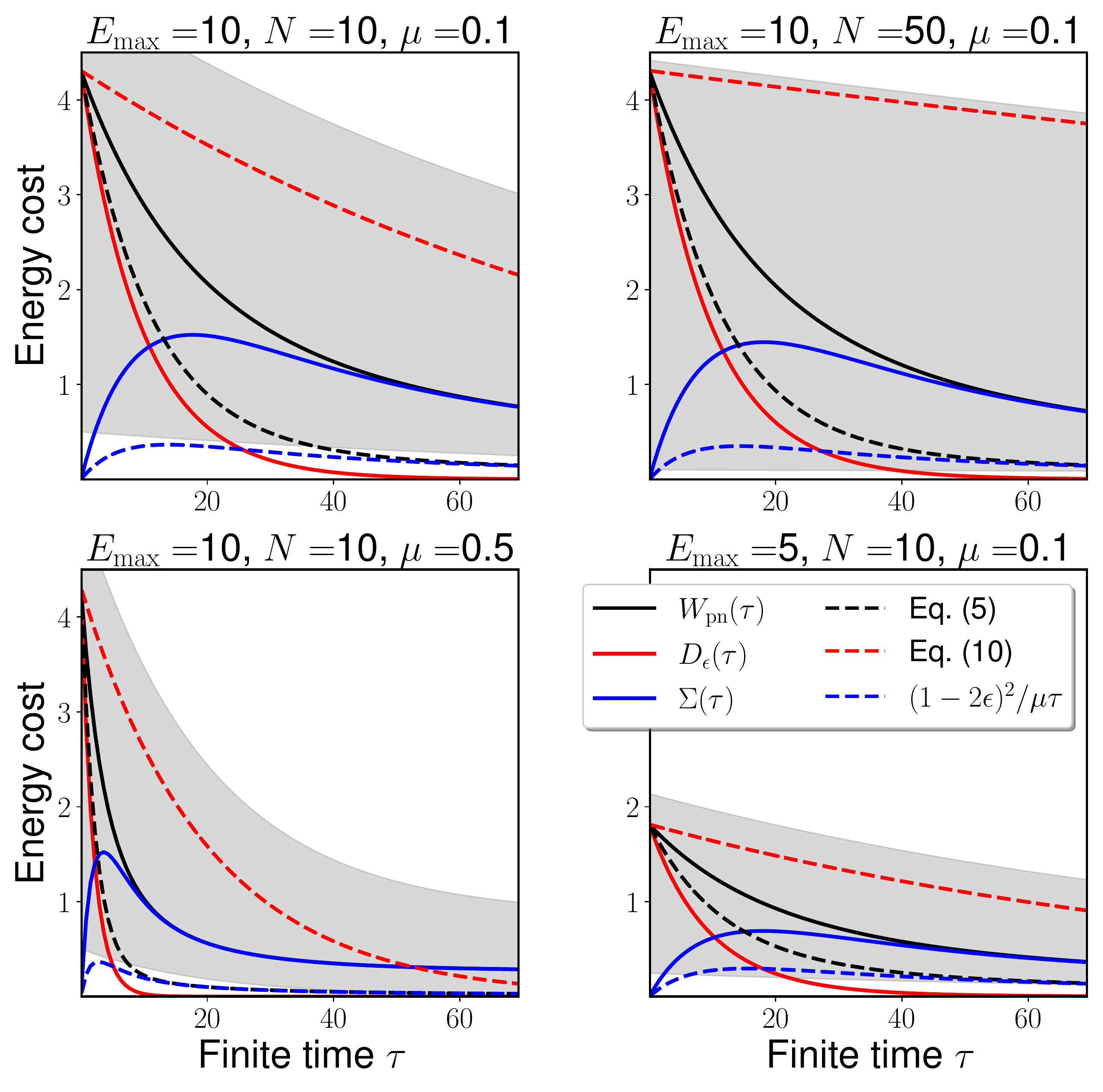}
\caption{\label{fig:Fig_fixE_comparison}%
Performance of the discrete-shifting protocol for different parameters when the maximal energy $E_{\max}$ is fixed ($\beta=1$). The shaded area is restricted by Eqs.~\eqref{eq:const-shifting-L} and~\eqref{eq:const-shifting-U}.}
\end{figure}

We consider protocols with fixed $E_{\max}$ and discrete shifting.
As $E_{\max}$ determines $\gamma(\tau)$, it further restricts the reset error by $\epsilon=P_1(\tau)\geqslant\gamma_1(\tau)$.
In Fig.~\ref{fig:Fig_fixE_EmaxVsEps} we show the dependence of the work penalty $W_{\rm pn}(\tau)$ and reset error $\epsilon$ on the finite time $\tau$ and the maximal energy $E_{\max}$.
When the finite time increases, both $W_{\rm pn}(\tau)$ and $\epsilon$ decrease.
For a given finite time $\tau$, $W_{\rm pn}(\tau)$ increases when $E_{\max}$ increases or when $\epsilon$ decreases.
Moreover, the bound in Eq.~\eqref{eq:work-penalty-bound} restricts $W_{\rm pn}(\tau)$ closely, and the tightness becomes better for small finite times.
This feature is mainly due to the tightness of the speed limit in Eq.~\eqref{eq:work-penalty-bound}, as shown in Fig.~\ref{fig:Fig_fixE_comparison}.
Furthermore, one might guess that the work penalty would explode when $\tau$ goes to zero, seen from Eq.~\eqref{eq:work-penalty-bound}. However, this is not the case here. As we have fixed $E_{\max}$, when $\tau\rightarrow0$, $E_1$ is almost instantly shifted from $0$ to $E_{\max}$, such that $\epsilon\rightarrow1/2$ and $W_{\rm pn}\rightarrow E_{\max}/2 - W_{\rm qs}(E_{\max})$.

In addition, we provide a lower bound for $\epsilon$ with derivations in Appendix~\ref{subsec:app-const-shifting-p1-lower}:
\begin{equation}\label{eq:const-shifting-lower-p1}
\epsilon\geqslant\gamma_{1}^{N}+e^{-\mu\tau/N}\gamma_{1}^{N}\left(1-\gamma_{1}^{N}\right)\left(1-e^{-\beta{\cal E}}\right),
\end{equation}
where $\gamma_1^N=1/(1+e^{\beta E_{\max}})$. This bound gives a minimal reset error achieved by the  discrete-shifting protocol. Indeed, as the bit system is driven by a non-equilibrium protocol, the final $P_1(\tau)$ is a bit larger than~$\gamma_1(\tau)$. 

The restrictions of different bounds on the work penalty, the relative entropy, and the entropy production under different parameters are shown in Fig.~\ref{fig:Fig_fixE_comparison}.
The $W_{\rm pn}(\tau)$ (solid black) and $D_{\epsilon}(\tau)$ (solid red) generally decrease as $\tau$ increases, while $\Sigma(\tau)$ (solid blue) is almost zero for short times as the system state is not changed too much for the small-time drive.
Recall that $D_{\epsilon}(\tau)$ is upper bounded by Eq.~\eqref{eq:const-shifting-relent-upper-discrete} (dashed red line) while $\Sigma(\tau)$ is lower bounded by the second term in Eq.~\eqref{eq:work-penalty-bound} (dashed blue line).
As a result, $W_{\rm pn}(\tau)$ is mainly contributed by $D_{\epsilon}(\tau)$ for short times and by $\Sigma(\tau)$ for long times, and consequently the work penalty initially drops exponentially and then drops inverse-linearly.
Moreover, due to the separation between $\Sigma(\tau)$ and the speed limit bound, the bound in Eq.~\eqref{eq:work-penalty-bound} becomes less tight for long times.
Furthermore, $W_{\rm pn}(\tau)$ decreases when $N$ increases as less heat is dissipated into the bath (and this change is slight in the figure because of the chosen parameters).
When $\mu$ increases, the system thermalizes faster such that the energy shift in each step costs less work and therefore $W_{\rm pn}(\tau)$ becomes less.
If one decreases $E_{\max}$, $W_{\rm pn}(\tau)$ will decrease significantly but the reset error will be large, which can also be seen from Fig.~\ref{fig:Fig_fixE_EmaxVsEps}.
As for the alternative bounds in Eqs.~\eqref{eq:const-shifting-L} and~\eqref{eq:const-shifting-U}, we can see that the upper bound in Eq.~\eqref{eq:const-shifting-U} restricts the exponential scaling of $W_{\rm pn}(\tau)$ generally while the lower bound in Eq.~\eqref{eq:const-shifting-L} performs better than Eq.~\eqref{eq:work-penalty-bound} only when both $N$ and $\tau$ are small.

\subsection{Case of fixed \texorpdfstring{$\epsilon$}{epsilon} }\label{subsec:const-shifting-fixed-eps}

\begin{figure}[htp]
\centering
\includegraphics[width=.9\columnwidth]{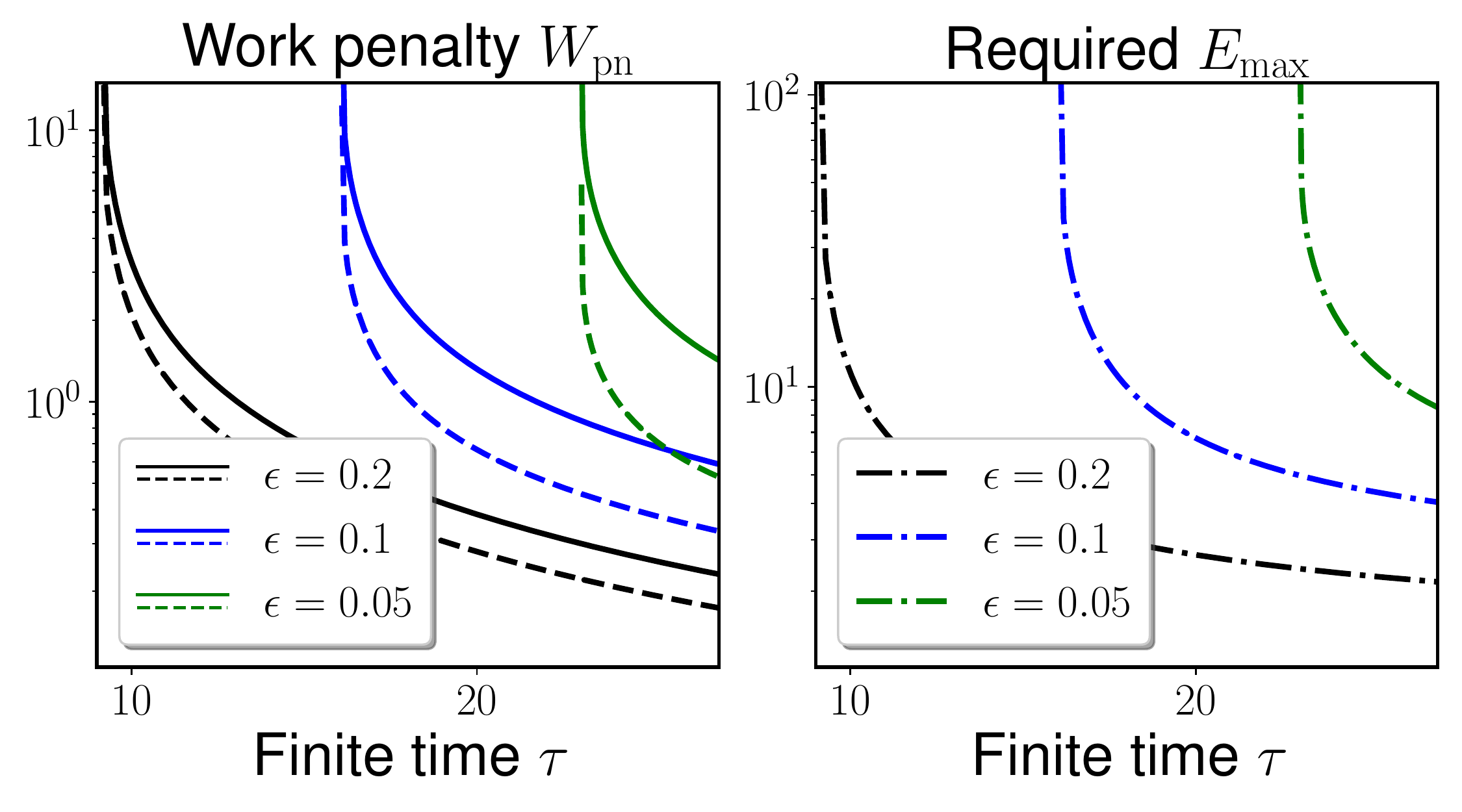}
\caption{\label{fig:Fig_fixP_EpsVsEmax}%
The discrete-shifting protocol with fixed reset error $\epsilon$. The solid lines, dashed lines and dashdotted lines represent the work penalty $W_{\rm pn}(\tau)$, the work penalty bound in Eq.~\eqref{eq:work-penalty-bound} and the required maximal energy level $E_{\max}$, respectively. $N=10$, $\mu=0.1$ and $\beta=1$ have been chosen for the simulation.}
\end{figure}

The dependence of the work penalty $W_{\rm pn}(\tau)$ and the required maximal energy level $E_{\max}$ on the finite time $\tau$ and the reset error $\epsilon$ is shown in Fig.~\ref{fig:Fig_fixP_EpsVsEmax}.
It can be seen that the bound of Eq.~\eqref{eq:work-penalty-bound} (dashed lines) restricts $W_{\rm pn}(\tau)$ (solid lines) closely.
For a finite time $\tau>\tau_{\epsilon}$, both $W_{\rm pn}(\tau)$ and the required $E_{\max}$ increase when $\epsilon$ decreases.
We derive in Appendix~\ref{subsec:app-const-shifting-emax-bds} that the required $E_{\max}$ satisfies
\begin{equation}\label{eq:const-shifting-emax-bounds}
\ln\frac{1-\epsilon}{\epsilon}\leqslant\beta E_{\max}\leqslant\ln\frac{1-\epsilon-\sqrt[N]{2\epsilon}/2}{\epsilon-\sqrt[N]{2\epsilon}/2}.
\end{equation}
Meanwhile, from the logarithmic scale in the figure, it can be seen that $W_{\rm pn}(\tau)$ drops faster and slower than the exponential scaling for short and long times, respectively.
We remark that for a fixed $\epsilon$, there exists a minimal finite time below which any reset protocol would fail to achieve $\epsilon$~\footnote{This minimal time $\tau_0$ is realized by the following protocol: $E_1$ is shifted to infinity at time $0$, then the system thermalises to the final state with $P_1(\tau_0)=\epsilon$.}.
\begin{figure}[htp]
\centering
\includegraphics[width=1.0\columnwidth]{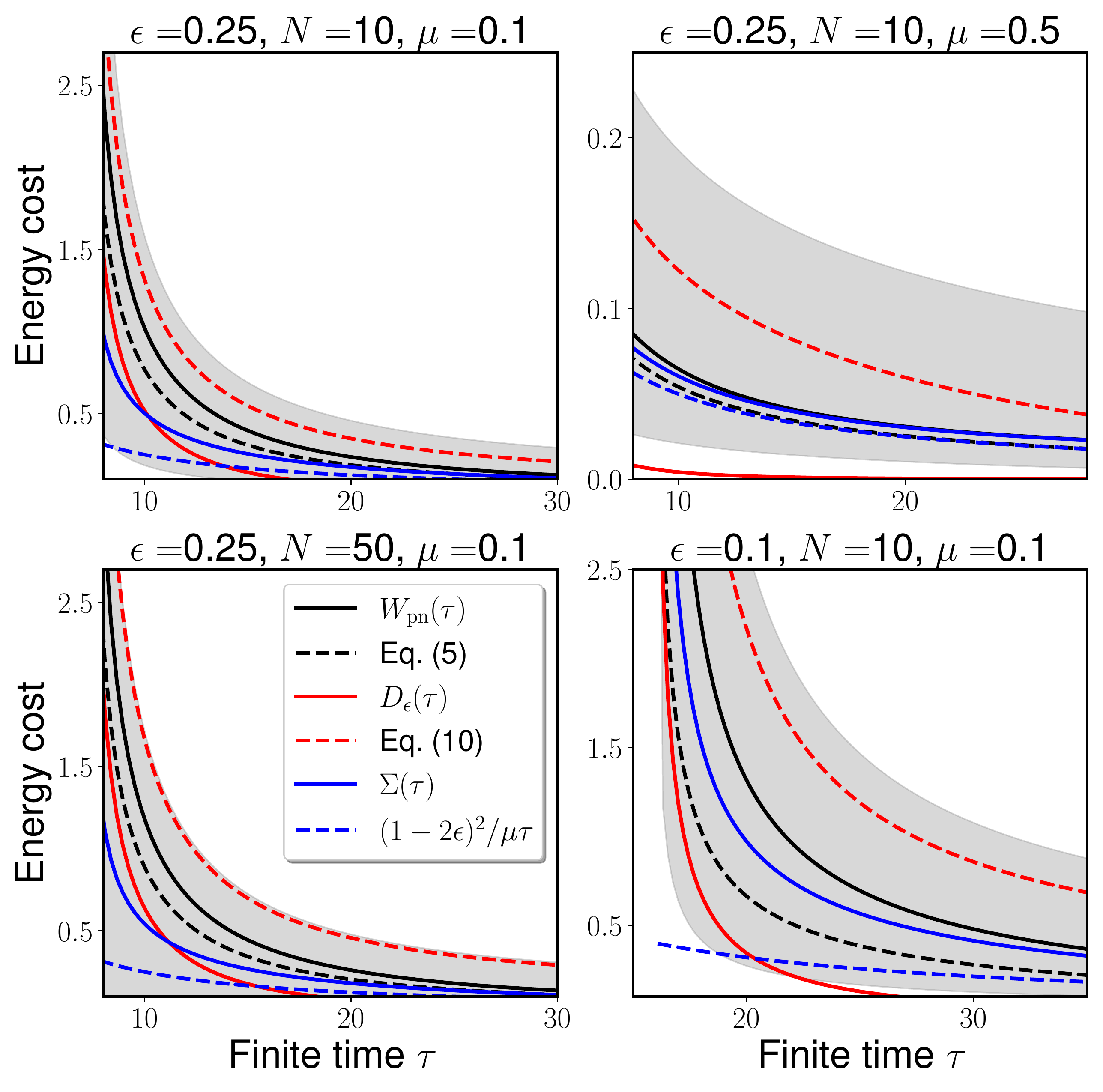}
\caption{\label{fig:Fig_fixP_comparison}%
Performance of the discrete-shifting protocol for different parameters when the reset error $\epsilon$ is fixed ($\beta=1$). The shaded area is restricted by Eqs.~\eqref{eq:const-shifting-L} and~\eqref{eq:const-shifting-U}.}
\end{figure}

To study how the bound in Eq.~\eqref{eq:work-penalty-bound} restricts $W_{\rm pn}(\tau)$ for different protocol parameters, we show in Fig.~\ref{fig:Fig_fixP_comparison} the work penalty $W_{\rm pn}(\tau)$, relative entropy $D_\epsilon(\tau)$ and entropy production $\Sigma(\tau)$ for given reset error $\epsilon$, shift number $N$, and swap rate $\mu$.
From the figure, we can see that $W_{\rm pn}(\tau)$ (black solid) is closely restricted by Eq.~\eqref{eq:work-penalty-bound} (black dashed).
As $W_{\rm pn}(\tau)$ is upper bounded by the right hand side of Eq.~\eqref{eq:const-shifting-relent-upper-discrete} for short times, we can conclude that $W_{\rm pn}(\tau)$ drops exponentially and then almost inverse linearly.
In addition, when $\epsilon$ is sufficiently small, $D_{\epsilon}(\tau)$ may be less than $\Sigma(\tau)$, such that $W_{\rm pn}(\tau)$ will mostly have an inverse-linear time scaling, as shown in the last panel of the figure.
We also observe that $W_{\rm pn}(\tau)$ becomes larger when $N$ increases, as in this case, the time interval of each thermalization step becomes small such that one requires a higher energy shift to drive the system to the desired state with reset error $\epsilon$.
Similarly, when $\mu$ increases, $W_{\rm pn}(\tau)$ decreases, as the system is better thermalized for each thermalization step such that the energy shift costs less energy. Thus also in this case the the work penalty decreases exponentially in $\tau$ for small $\tau$ and inverse linearly in large $\tau$.

\section{Continuous driving protocol}\label{sec:general-reset-protocol}

For continuous driving, the protocol can be described by a function of $E_1(t)$ with $E_1(0)=0$ and $E_1(\tau)=E_{\max}$.
Here, we investigate if the time scaling discussed in the discrete case can be extended to the continuous case.
As the entropy production can be lower bounded via the speed limit approach, i.e.\ $\Sigma(\tau)\geqslant (1-2\epsilon)^2/\mu\tau$,
we can immediately conclude that the work penalty cannot escape the inverse-linear scaling when $\Sigma$ dominates.
In contrast, if $D_{\epsilon}(\tau)$ dominates the work penalty, a different scaling may occur.
Unfortunately, the upper bound of relative entropy in Eq.~\eqref{eq:const-shifting-relent-upper-discrete} does not apply to the continuous case.
Instead, we analytically prove that there always exists a sufficiently small finite time before $D_{\epsilon}(\tau)$ drops exponentially.
We specifically consider a finite time $\tau\leqslant-\mu^{-1}\ln\epsilon$.
Then, $D_{\epsilon}(\tau)$ can be proved to be bounded by
\begin{equation}\label{eq:continuous-rel-entropy-upper-bound}
D_{\epsilon}\left(\tau\right)\leqslant e^{-\mu\tau}\ln\left(1+e^{\beta E_{\max}}\right).
\end{equation}
In a more general sense (for general $\tau$), we can show that $D_{\epsilon}(\tau)$ is upper bounded by
\begin{align}
D_{\epsilon}\left(\tau\right) \leqslant & e^{-\mu\tau}D\left[\gamma(0)\|\gamma(\tau)\right]\nonumber\\
&+ \left(1- e^{-\mu\tau}\right)D\left[\gamma(0^{+})\|\gamma(\tau)\right].\label{eq:general-rel-entropy-upper-bound}
\end{align}
We provide the proofs of the above two inequalities in Appendix~\ref{subsec:app-relat-entropy-bound-continuous}.

To numerically investigate the case of the continuous protocol, we particularly consider a special protocol where the driving is linear but an initial-time energy jump exits.
Indeed, linear driving is arguably the easiest control in practice while the initial energy jump has been employed in many optimal protocols (see, e.g., Refs.~\cite{SchmiedlS07, ZulkowskiD14, ProesmansEB20, ProesmansEB20pre}).
To further simplify the analysis, let the energy jump be $\Delta/\beta$ and the speed of driving be $\mu/\beta$, i.e.,
\begin{equation}\label{eq:continuous-shifting-protocol}
E_1(t) = (\mu t + \Delta) / \beta,\qquad 0<t\leqslant \tau.
\end{equation}
The convenience of this protocol is that the bit evolution can be analytically solved, based on partial swap model (see Eq.~\eqref{eq:partial-swap-model-solution}), as
\begin{equation}\label{eq:continuous-shifting-P1t}
P_1(t) = e^{-\mu t}\left[\frac{1}{2}+ e^{-\Delta}\ln\frac{1+e^{\mu t+\Delta}}{1+e^{\Delta}}\right].
\end{equation}
From this equation, $W_{\rm pn}(\tau)$, $D_{\epsilon}(\tau)$, $\Sigma(\tau)$ and associated bounds can be calculated.
Similar to the case of discrete-shifting protocol, we study this continuous protocol for cases of fixed maximal energy level and fixed reset error, as shown from Figs.~\ref{fig:Fig_continuous_fixE_EmaxVsEps} to~\ref{fig:Fig_continuous_fixP_comparison}.

\subsection{Case of fixed \texorpdfstring{$E_{\max}$}{Emax} }\label{subsec:continuous-fix-emas}

\begin{figure}[htp]
\centering
\includegraphics[width=0.9\columnwidth]{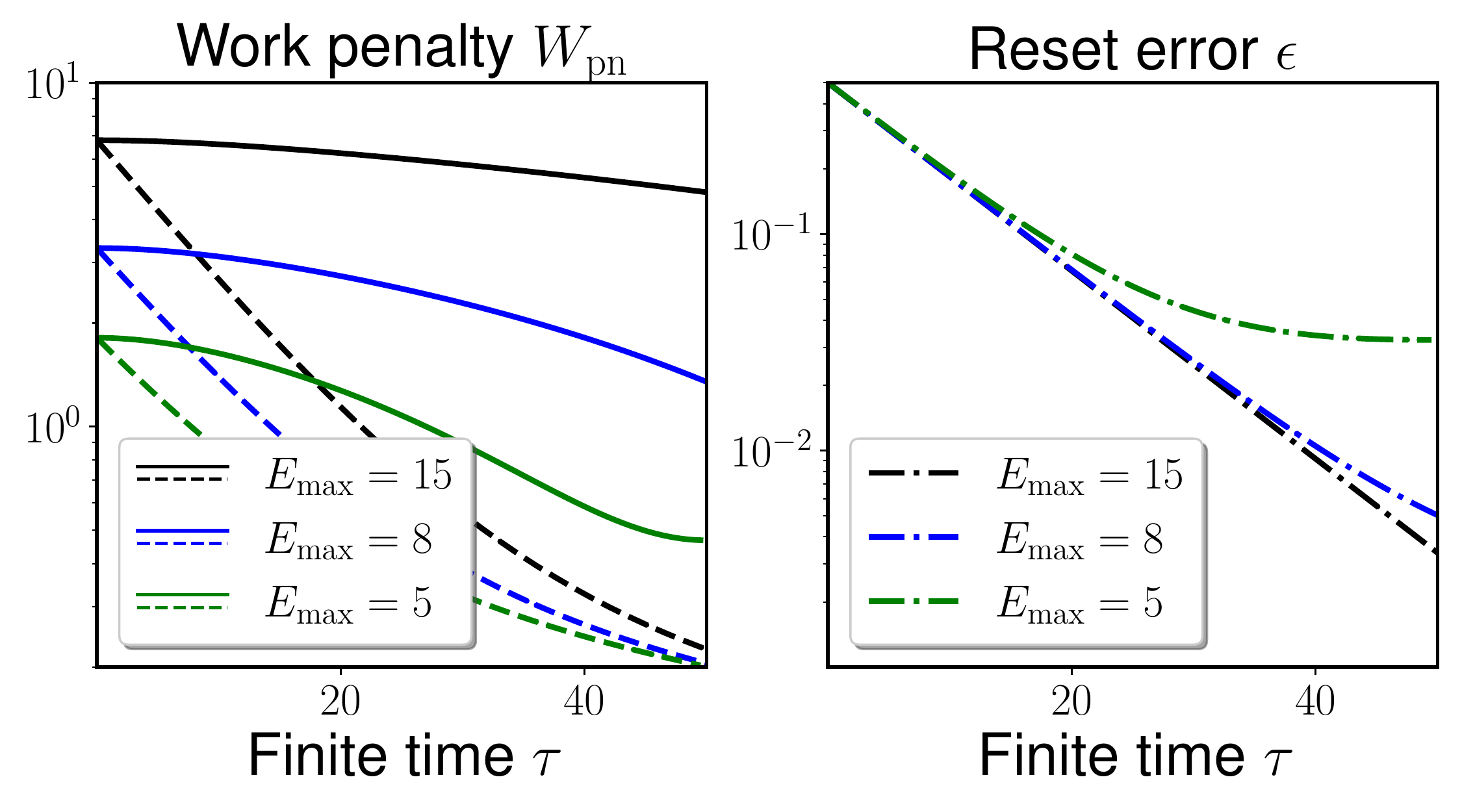}
\caption{\label{fig:Fig_continuous_fixE_EmaxVsEps}%
The continuous driving protocol as described in Eq.~\eqref{eq:continuous-shifting-protocol} with fixed maximal energy level $E_1(\tau)=E_{\max}$. The solid lines, dashed lines and dashdotted lines represent the work penalty $W_{\rm pn}(\tau)$, the work penalty bound in Eq.~\eqref{eq:work-penalty-bound} and the reset error $\epsilon$, respectively. $\mu=0.1$ and $\beta=1$ have been chosen for the simulation.}
\end{figure}

\begin{figure}[htp]
\centering
\includegraphics[width=1.0\columnwidth]{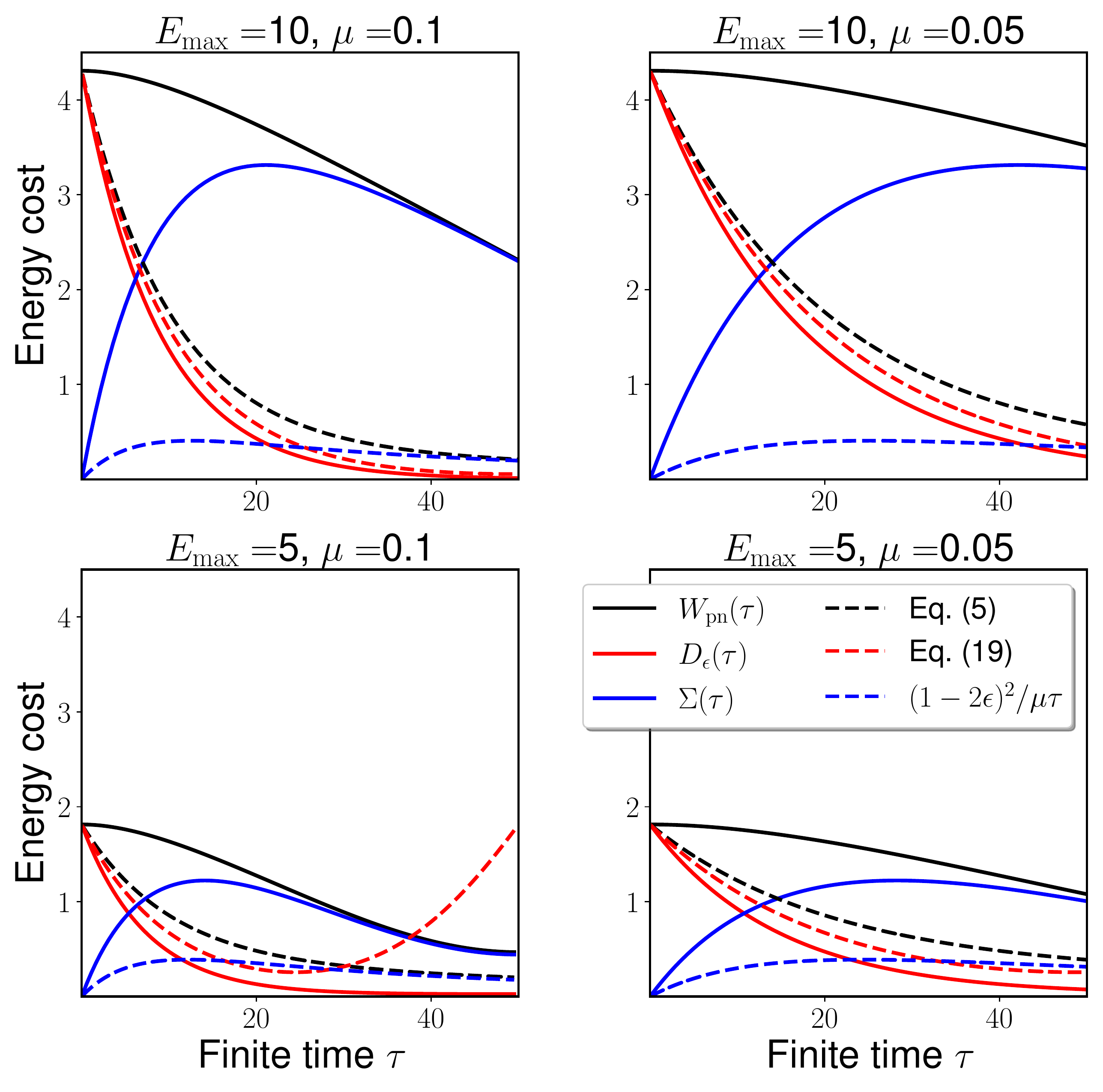}
\caption{\label{fig:Fig_continuous_fixE_comparison}%
Performance of the continuous driving protocol as described in Eq.~\eqref{eq:continuous-shifting-protocol} for different parameters, when the maximal energy $E_{\max}$ is fixed ($\beta=1$).}
\end{figure}

For the case of fixed maximal energy level, i.e., $E_1(t=\tau)=E_{\max}$, we show in Figs.~\ref{fig:Fig_continuous_fixE_EmaxVsEps} and~\ref{fig:Fig_continuous_fixE_comparison} that $W_{\rm pn}(\tau)$ decreases as the finite time $\tau$ increases, while the work penalty bound in Eq.~\eqref{eq:work-penalty-bound} restricts $W_{\rm pn}(\tau)$ closely.
Similar to the discrete case, $W_{\rm pn}(\tau)$  approaches $E_{\max}/2$ when $\tau$ is almost zero.
From the slope of the solid and dash-dotted lines in Fig.~\ref{fig:Fig_continuous_fixE_EmaxVsEps}, we can see that both the work penalty and the reset error drops almost exponentially for small times of $\tau$.
Particularly, when $E_{\max}$ increases, for a given finite time $\tau$, the work penalty increases while the reset error decreases.
In Fig.~\ref{fig:Fig_continuous_fixE_comparison}, however, the exponential decay of $W_{\rm pn}(\tau)$ is not obvious.
This is due to the fact $\Sigma(\tau)$ increases quickly, while $D_{\epsilon}(\tau)$ is upper bounded by Eq.~\eqref{eq:general-rel-entropy-upper-bound} closely in small times.
Furthermore, if one increases $\mu$ such that the system can thermalize better, $W_{\rm pn}(\tau)$, $D_{\epsilon}(\tau)$ and $\Sigma(\tau)$ become smaller, as shown in Fig.~\ref{fig:Fig_continuous_fixE_comparison}.

\subsection{Case of fixed \texorpdfstring{$\epsilon$}{epsilon} }\label{subsec:continuous-fix-eps}

\begin{figure}[htp]
\centering
\includegraphics[width=0.9\columnwidth]{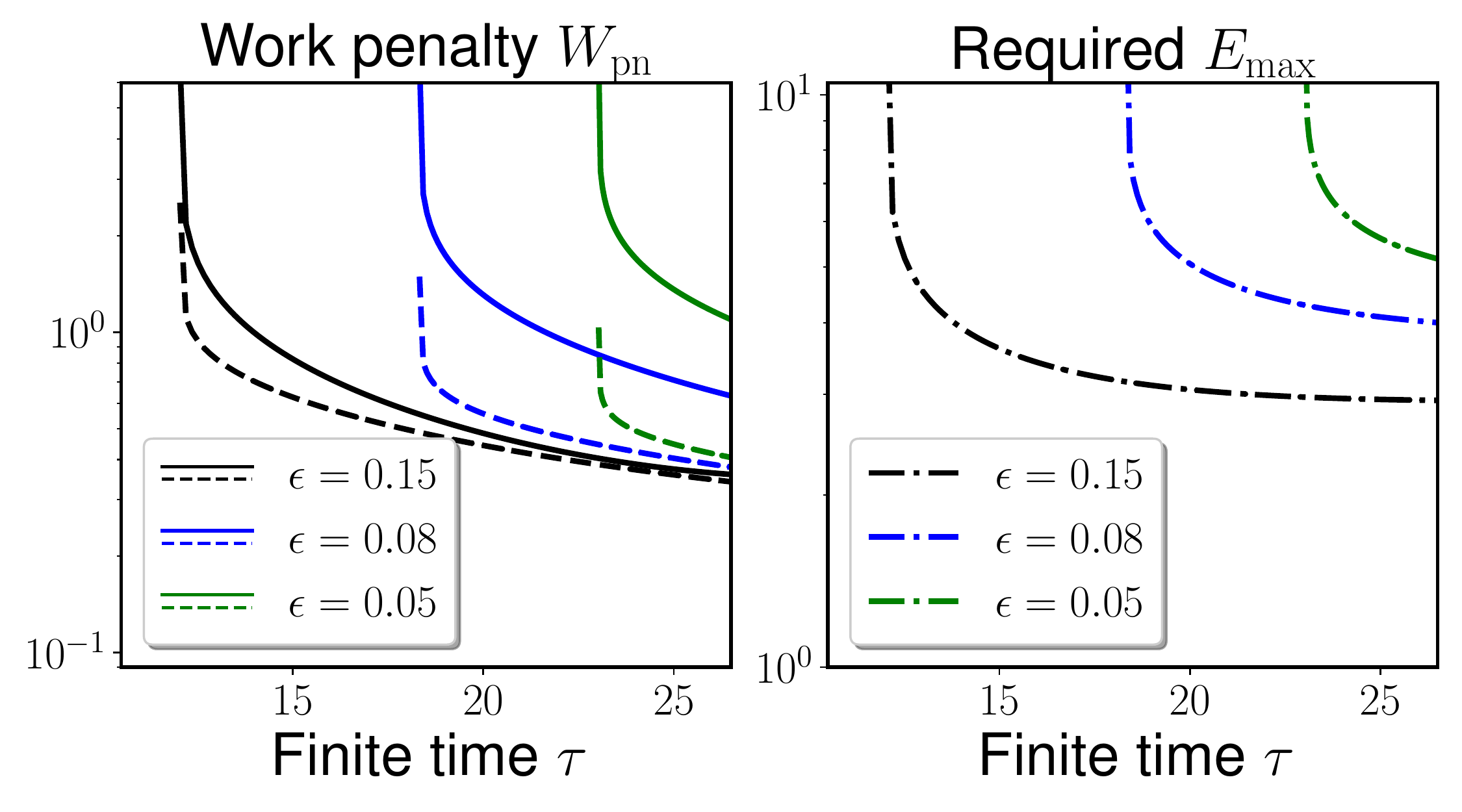}
\caption{\label{fig:Fig_continuous_fixP_EpsVsEmax}%
The continuous driving protocol as described in Eq.~\eqref{eq:continuous-shifting-protocol} with fixed reset error $P_1(\tau)=\epsilon$. The solid lines, dashed lines and dashdotted lines represent the work penalty $W_{\rm pn}(\tau)$, the work penalty bound in Eq.~\eqref{eq:work-penalty-bound} and the required maximal energy level $E_{\max}$, respectively. $\mu=0.1$ and $\beta=1$ have been chosen for the simulation.}
\end{figure}

\begin{figure}[htb]
\centering
\includegraphics[width=1.0\columnwidth]{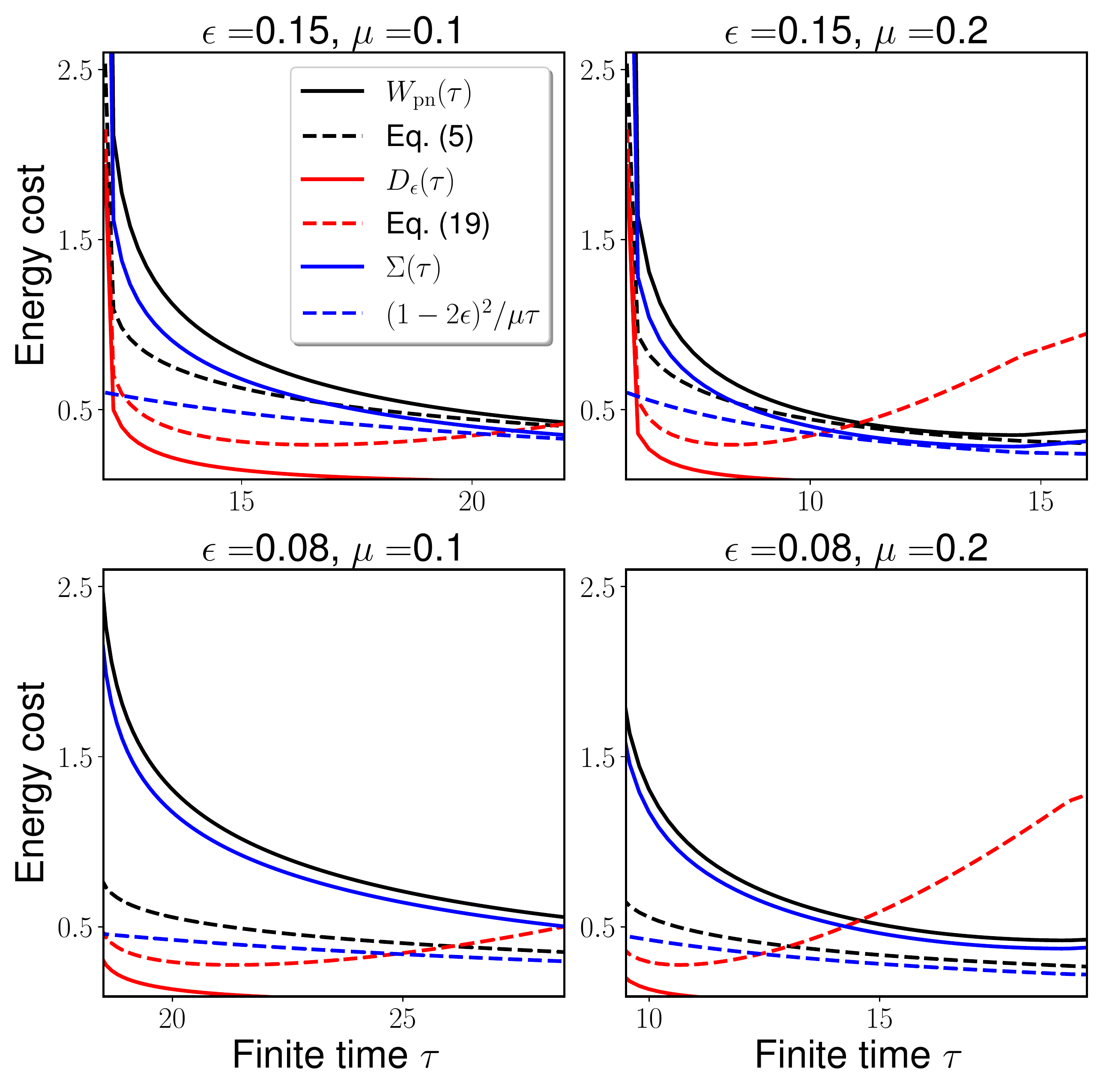}
\caption{\label{fig:Fig_continuous_fixP_comparison}%
Performance of the continuous driving protocol as described in Eq.~\eqref{eq:continuous-shifting-protocol} for different parameters, when the reset error $\epsilon$ is fixed ($\beta=1$).}
\end{figure}

For the case of fixed reset error, i.e., $P_1(\tau)=\epsilon$, we show in Figs.~\ref{fig:Fig_continuous_fixP_EpsVsEmax} and~\ref{fig:Fig_continuous_fixP_comparison} that $W_{\rm pn}(\tau)$ can be restricted closely by the bound in Eq.~\eqref{eq:work-penalty-bound}, implying that $W_{\rm pn}(\tau)$ cannot escape the inverse-linear scaling.
There is also a minimal finite time below which any reset protocol would fail to achieve $\epsilon$~\cite{Note1}.
Moreover, $W_{\rm pn}(\tau)$ decreases faster than exponential drops for short times, as shown by the slopes of the solid lines in the left panel of Fig.~\ref{fig:Fig_continuous_fixP_EpsVsEmax}.
Consequently, the required $E_{\max}$ to obtain a final $\epsilon$ also drops sharply.
If one requires the protocol to reset the bit better, i,e., demanding a smaller $\epsilon$, one then needs to increase $E_{\max}$ and apply more work if the finite time $\tau$ is small.
For the trade-off contribution between $D_{\epsilon}(\tau)$ and $\Sigma(\tau)$ to $W_{\rm pn}(\tau)$, we show in Fig.~\ref{fig:Fig_continuous_fixP_comparison} that $D_{\epsilon}(\tau)$ drops extremely fast such that $W_{\rm pn}(\tau)$ is mainly due to $\Sigma(\tau)$ for most finite times.

\section{Conclusion}\label{sec:conclustion}

In this paper, by considering an effective two-level system, we have investigated the scaling of the work penalty with the protocol parameters, including the time allowed, the bit reset error allowed and the final energy gap. We derived results for two specific protocols: the  discrete-shifting protocol and the continuous driving protocol.
We have shown that for both protocols, the work penalty decays exponentially when considering a sufficiently small finite time and decays inverse-linearly when the finite time is large.
This is due to the trade-off in contribution between the relative entropy and the entropy production to the work penalty.
Above all, our work shows in detail how the work penalty of finite-time bit reset depends on the protocol requirements. The result additionally helps the design of reset protocols to balance the energy cost and reset errors. 
Particularly, for a certain hardware where the maximal energy gap between the bit-value states is fixed and for a certain demanded reset error, our result can be used to select suitable reset protocols. In scenarios where the time is constrained, one can strike a balance between the time and reset error by setting the parameters such that the work penalty falls beyond the exponential region but not too far into the inverse-linear region.


There are related results that are based on information geometry, and provide efficient tools to study the optimal protocols such that the entropy production can be lower bounded.
Particularly, 
an information geometric lower bound with $1/\tau$ scaling was given in Ref.~\cite{VanVuH2021}. To evaluate the (Wasserstein) distance in the bound one can use either a bound in terms of total variation distance and a measure of the strength of the time evolution, or numerical methods for evaluating geodesics.
Ref.~\cite{NakazatoI2021} 
employs similar methods to create tight bounds 
under a Fokker-Planck master equation.
Ref.~\cite{Dechant2022} shows how to derive analogous results for discrete state spaces.
The information geometry approach was moreover employed to derive 
an expression for the optimal power of heat engines operating between two temperatures in Ref.~\cite{FuTCG2021}. 
These information geometry type results are, as described, e.g., in Ref.~\cite{VanVuH2021} 
closely related to the speed-limit on entropy production employed here.
They offer a complementary route to show $1/\tau$ scaling in the long time regime and tools to determine optimal values numerically. It would be interesting to apply the information geometry approach to analyse the exponential scaling in the short-time regime.

\begin{acknowledgments}
We acknowledge valuable discussions with Yingqiu Mao, Yutong Luo. Y.Z.Z.\ and O.D.\ acknowledge support from the National Natural Science Foundation of China (Grant No.~12050410246,~12005091).
D.E.\ acknowledges support from the Swiss National Science Foundation (Grant~No.~P2SKP2\_18406).
K.M.\ acknowledges support from the Australian Academy of Technology and Engineering via the 2018 Australia China Young Scientists Exchange Program and the 2019 Next Step Initiative.
Y.Z.Z.\ also acknowledges China Postdoctoral Science Foundation (No.~2020M671856).
\end{acknowledgments}

\appendix

\section{Derivations in the main text}\label{sec:app-derivations}

\subsection{Proof of \texorpdfstring{Eq.~\eqref{eq:relat-entropy-lower-bound}}{Eq. (8)}}\label{subsec:app-relat-entropy-bound}

We firstly show that the partial swap model in Eq.~\eqref{eq:partial-swap-model} can be solved as
\begin{equation}\label{eq:partial-swap-model-solution}
P_{1}\left(t\right)=e^{-\mu t}\left[P_1(0)+\mu\int_{0}^{t}ds e^{\mu s}\gamma_{1}\left(s\right)\right].
\end{equation}
An efficient reset protocol requires $E_1(s)\leqslant E_1(t)$ for any two times $s\leqslant t$.
This leads to $\gamma_1(t)=1/(1+\exp(\beta E_1(t))$  monotonically decreasing with time such that $\gamma_1(s) \geqslant \gamma_1(t)$.
Substituting this relation into Eq.~\eqref{eq:partial-swap-model-solution} and applying the initial condition $P_1(0)=\gamma_1(0)$, we have
\begin{equation}\label{eq:P1-lower-bound}
P_1(\tau) \geqslant 
 e^{-\mu\tau}\gamma_1(0)+\left(1-e^{-\mu \tau}\right)\gamma_1(\tau).
\end{equation}

Meanwhile, $D[P\|\gamma]$ is monotonically increasing with $P_1$, which can be verified by
\begin{equation}\label{eq:tmp2}
\frac{\partial}{\partial P_1}D\left[P\|\gamma\right] 
=\ln\frac{(1-\gamma_1)P_1}{(1-P_1)\gamma_1}\geqslant 0,
\end{equation}
where we have used the fact that $P_1\geqslant\gamma_1\geqslant 0$.
Using the inequality of Eq.~\eqref{eq:P1-lower-bound}, and for convenience denoting $q=\exp(-\mu\tau)$ and ${\tilde \gamma} = q\gamma(0)+\left(1-q\right)\gamma(\tau)$, we can bound the relative entropy as
\begin{align}
 & D\left[P\left(\tau\right)\|\gamma\left(\tau\right)\right]\\
\geqslant & D\left[\tilde{\gamma}\|\gamma\left(\tau\right)\right]=\sum_{a}\tilde{\gamma}_{a}\ln\frac{\tilde{\gamma}_{a}}{\gamma_{a}\left(\tau\right)}\\
= & \sum_{a}\left[q\gamma_{a}\left(0\right)+\left(1-q\right)\gamma_{a}\left(\tau\right)\right]\ln\frac{\tilde{\gamma}_{a}}{\gamma_{a}\left(\tau\right)}\\
= & q\sum_{a}\gamma_{a}\left(0\right)\left[\ln\frac{\tilde{\gamma}_{a}}{\gamma_{a}\left(0\right)}+\ln\frac{\gamma_{a}\left(0\right)}{\gamma_{a}\left(\tau\right)}\right]\nonumber\\
 & +\left(1-q\right)\sum_{a}\gamma_{a}\left(\tau\right)\ln\frac{\tilde{\gamma}_{a}}{\gamma_{a}\left(\tau\right)}\\
= & -qD\left[\gamma\left(0\right)\|\tilde{\gamma}\right]+qD\left[\gamma\left(0\right)\|\gamma\left(\tau\right)\right]\nonumber\\
 & -\left(1-q\right)D\left[\gamma\left(\tau\right)\|\tilde{\gamma}\right]\\
\geqslant & q^{2}D\left[\gamma\left(0\right)\|\gamma\left(\tau\right)\right]\nonumber\\
 & -q\left(1-q\right)D\left[\gamma\left(\tau\right)\|\gamma\left(0\right)\right]
\end{align}
where we have used the convexity of relative entropy. 
Since
\begin{align}
D\left[\gamma\left(\tau\right)\|\gamma\left(0\right)\right] & =\sum_{a=0,1}\gamma_{a}\left(\tau\right)\ln\frac{\gamma_{a}\left(\tau\right)}{1/2}\\
 & =\sum_{a=0,1}\gamma_{a}\left(\tau\right)\ln\frac{2e^{-\beta E_{a}\left(\tau\right)}}{Z\left(\tau\right)}\\
 & =\ln\frac{2}{Z\left(\tau\right)}-\sum_{a}\gamma_{a}\beta E_{a}\left(\tau\right)\\
 & =\beta\left[W_{{\rm qs}}\left(\tau\right)-\epsilon E_{\max}\right]
\end{align}
\begin{align}
D\left[\gamma\left(0\right)\|\gamma\left(\tau\right)\right]&=\sum_{a=0,1}\frac{1}{2}\ln\frac{1/2}{\gamma_{a}\left(\tau\right)}\\
&=-\ln2+\frac{1}{2}\ln\frac{1}{\gamma_{0}\left(\tau\right)\gamma_{1}\left(\tau\right)}\\
&=\frac{1}{2}\ln\left(\frac{1+\cosh\left(\beta E\right)}{2}\right)
\end{align}
Finally, we have
\begin{align}
& D\left[P\left(\tau\right)\|\gamma\left(\tau\right)\right]\\
\geqslant & \max\left\{0, e^{-2\mu t}G_{1}-e^{-\mu t}\left(1-e^{-\mu t}\right)G_{2}\right\} ,
\end{align}
where $G_{1}=\ln\left[\left(1+\cosh\left(\beta E_{\max}\right)\right)/2\right]/2$
and $G_{2}=\beta\left(W_{{\rm pn}}\left(\tau\right)-\epsilon E_{\max}\right)$.

\subsection{Proof of \texorpdfstring{Eq.~\eqref{eq:const-shifting-sigma-dominates}}{Eq. (11)}}\label{subsec:app-const-shifting-sigma-dominates}

In the discrete-shifting protocol, the work penalty can be written as $\beta W_{{\rm pn}}(\tau) =D\left[P^{N}\|\gamma^{N}\right]+\Sigma(\tau)$, where $\Sigma(\tau)=\sum_{k=1}^{N}\Sigma^{k}$ is contributed by the entropy production of each thermalization step, i.e.
\begin{equation}\label{eq:tmp56}
\Sigma^{k}=D\left[P^{k-1}\|\gamma^{k}\right]-D\left[P^{k}\|\gamma^{k}\right].
\end{equation}
Similarly, we also introduce the relative entropy changes for each step of energy shift and thermalization such that $D\left[P^{N}\|\gamma^{N}\right] = \sum_{k=1}^{N}D^{k}$ with
\begin{equation}\label{eq:tmp60}
D^{k} =D\left[P^{k}\|\gamma^{k}\right]-D\left[P^{k-1}\|\gamma^{k-1}\right].
\end{equation}

We consider a sufficient but not necessary condition for $D\leqslant\Sigma$, i.e.\ when $D^{k}\leqslant\Sigma^{k}$ for all $k$.
This condition is equivalent to
\begin{equation}\label{eq:tmp66}
2D\left[P^{k}\|\gamma^{k}\right]\leqslant D\left[P^{k-1}\|\gamma^{k-1}\right]+D\left[P^{k-1}\|\gamma^{k}\right],
\end{equation}
for all $k$.
Note that in the supplementary materials of~\cite{ZhenEMD21} we have proved that $D\left[P^{k}\|\gamma^{k}\right]\leqslant e^{-2\mu\tau/N}D\left[P^{k-1}\|\gamma^{k}\right]$.
Then, to obtain Eq.~\eqref{eq:tmp66} we can require $ 2e^{-2\mu\tau/N}D\left[P^{k-1}\|\gamma^{k}\right]\leqslant D\left[P^{k-1}\|\gamma^{k-1}\right]+D\left[P^{k-1}\|\gamma^{k}\right]$, which is equivalent to
\begin{equation}\label{eq:tmp71}
2e^{-2\mu\tau/N}\leqslant\frac{D\left[P^{k-1}\|\gamma^{k-1}\right]}{D\left[P^{k-1}\|\gamma^{k}\right]}+1.
\end{equation}
This equation is satisfied if $2e^{-2\mu\tau/N}\leqslant1$, i.e., when $\tau\geqslant N\ln2/(2\mu)$.
Therefore, if $\tau$ satisfies this condition, we must have $D^k \leqslant \Sigma^k$ for each $k$ such that $W_{{\rm pn}}(\tau)$ is
dominated by $\Sigma(\tau)$.

\subsection{Proof of \texorpdfstring{Eq.~\eqref{eq:const-shifting-lower-p1}}{Eq. (14)}}\label{subsec:app-const-shifting-p1-lower}

From Eq.\eqref{eq:const-shifting-P1}, the reset error can be bounded by
\begin{align}
\epsilon & =P_{1}^{N}=e^{-\mu\tau/N}P_{1}^{N-1}+\left(1-e^{-\mu\tau/N}\right)\gamma_{1}^{N}\\
 & \geqslant\gamma_{1}^{N}+e^{-\mu\tau/N}\left(\gamma_{1}^{N-1}-\gamma_{1}^{N}\right),
\end{align}
where we have used $P_{1}^{N-1}\geqslant\gamma_{1}^{N-1}$.
Since $\gamma_{1}^{k}=1/\left(1+e^{\beta kE_{\max}/N}\right)$, we have
\begin{align}
\gamma_{1}^{N-1}-\gamma_{1}^{N} & =\frac{1}{1+e^{\beta E_{\max}}e^{-\beta{\cal E}}}-\frac{1}{1+e^{\beta E_{\max}}}\\
 & =\frac{e^{\beta E_{\max}}-e^{\beta E_{\max}}e^{-\beta{\cal E}}}{\left(1+e^{\beta E_{\max}}e^{-\beta{\cal E}}\right)\left(1+e^{\beta E_{\max}}\right)}\\
 & \geqslant\frac{e^{\beta E_{\max}}}{\left(1+e^{\beta E_{\max}}\right)^{2}}\left(1-e^{-\beta{\cal E}}\right)\\
 & =\gamma_{1}^{N}\left(1-\gamma_{1}^{N}\right)\left(1-e^{-\beta{\cal E}}\right),
\end{align}
where ${\cal E}=E_{\max}/N$.
Therefore,
\begin{equation}\label{eq:tmp68}
\epsilon\geqslant\gamma_{1}^{N}+e^{-\mu\tau/N}\gamma_{1}^{N}\left(1-\gamma_{1}^{N}\right)\left(1-e^{-\beta{\cal E}}\right)
\end{equation}

\subsection{Proof of \texorpdfstring{Eq.~\eqref{eq:const-shifting-emax-bounds}}{Eq. (15)}}\label{subsec:app-const-shifting-emax-bds}

For the case of fixed reset error $\epsilon,$ we have shown 
that $\mu\tau\geqslant-\ln\left(2\epsilon\right)$.
Together with Eq.~\eqref{eq:const-shifting-P1}, we can show that
\begin{align}
\epsilon & =e^{-\mu\tau/N}P_{1}^{N-1}+\left(1-e^{-\mu\tau/N}\right)\gamma_{1}^{N}\\
 & \leqslant\gamma_{1}^{N}+e^{-\mu\tau/N}\left(\frac{1}{2}-\gamma_{1}^{N}\right)\\
 & \leqslant\gamma_{1}^{N}+\sqrt[N]{2\epsilon}\left(\frac{1}{2}-\gamma_{1}^{N}\right),
\end{align}
where we have used $P_{1}^{N-1}\leqslant1/2$.
Then, we obtain
\begin{equation}\label{eq:tmp111}
\ln\frac{1-\epsilon-\sqrt[N]{2\epsilon}/2}{\epsilon-\sqrt[N]{2\epsilon}/2}\geqslant\beta E_{\max}.
\end{equation}
Moreover, using the fact that
\begin{equation}\label{eq:tmp115}
\epsilon\geqslant\gamma_{1}\left(\tau\right)=\frac{1}{1+e^{\beta E_{\max}}},
\end{equation}
we have
\begin{equation}\label{eq:tmp119}
\beta E_{\max}\geqslant\ln\frac{1-\epsilon}{\epsilon}.
\end{equation}

\subsection{Proof of \texorpdfstring{Eq.~\eqref{eq:continuous-rel-entropy-upper-bound}}{Eq. (16)} and \texorpdfstring{Eq.~\eqref{eq:general-rel-entropy-upper-bound}}{Eq. (17)}}\label{subsec:app-relat-entropy-bound-continuous}

For the reset error $\epsilon$, let
\begin{equation}\label{eq:tmp98}
\tau_{D}=\frac{1}{\mu}\ln\frac{1}{\epsilon}.
\end{equation}
We define a binary distribution $Q=[1-Q_{1},Q_{1}]$ with
\begin{equation}\label{eq:tmp102}
Q_{1}=e^{\mu\tau}\epsilon\leqslant1,\qquad\text{for }\tau\leqslant\tau_{D}.
\end{equation}
This leads to the relation
\begin{align}
P_{1}\left(\tau\right) & =\epsilon=e^{-\mu\tau}Q_{1}\\
 & \leqslant e^{-\mu\tau}Q_{1}+\left(1-e^{-\mu\tau}\right)\gamma_{1}\left(\tau\right),\\
& \qquad\text{for }\tau\leqslant\tau_{D}.
\end{align}
According to the monotonically increasing property of $D\left[P\|\gamma\right]$
with respective to $P_{1}$, we have Eq.~\eqref{eq:continuous-rel-entropy-upper-bound}, i.e.
\begin{align}
D_{\epsilon}\left(\tau\right) & =D\left[P\left(\tau\right)\|\gamma\left(\tau\right)\right]\\
 & \leqslant D\left[e^{-\mu\tau}Q+\left(1-e^{-\mu\tau}\right)\gamma\left(\tau\right)\|\gamma\left(\tau\right)\right]\\
 & \leqslant e^{-\mu\tau}D\left[Q\|\gamma\left(\tau\right)\right],
\end{align}
where the last inequality comes from the convexity of relative entropy.
Moreover, $D\left[Q\|\gamma\left(\tau\right)\right]$ can be trivially upper bounded by replacing $Q$ with deterministic distribution $[0,1]$, then
\begin{equation}
D\left[Q\|\gamma\left(\tau\right)\right] \leqslant \ln \frac{1}{\gamma_1(\tau)} = \ln \left(1+e^{\beta E_{\max}}\right)
\end{equation}

In the general case, we have that
\begin{equation}\label{eq:tmp159}
P_{1}\left(\tau\right)\leqslant e^{\mu \tau}P_{1}\left(0\right)+\left(1-e^{\mu \tau}\right)\gamma_{1}\left(0^{+}\right).
\end{equation}
Using the above procedure again, we have
\begin{align}
D_{\epsilon}(\tau) \leqslant & e^{-\mu\tau}D\left[P\left(0\right)\|\gamma\left(\tau\right)\right]\\
&+\left(1-e^{-\mu\tau}\right)D\left[\gamma\left(0^{+}\right)\|\gamma\left(\tau\right)\right]
\end{align}
such that Eq.~\eqref{eq:general-rel-entropy-upper-bound} is obtained.

\bibliography{Ref_bitreset}

\end{document}